\documentclass[1p]{elsarticle}

\PassOptionsToPackage{hyphens}{url}\usepackage[hyperindex,
linktocpage,
bookmarksnumbered,
plainpages=false,
pdfpagelabels,
colorlinks,
bookmarks,
linkcolor=black,
urlcolor=black,
citecolor=black,
pdfpagemode=UseOutlines,
hyperfootnotes = false,
breaklinks=true]{hyperref}

\usepackage[labelfont=bf,justification=raggedright,singlelinecheck=false]{caption}
\usepackage{float}
\usepackage{lineno,hyperref}
\usepackage{subfig}
\usepackage{color}
\usepackage{amsfonts}
\renewcommand{\vec}[1]{\ensuremath{\boldsymbol{#1}}} 
\graphicspath{{figures/}}
\usepackage[cmex10]{amsmath}

\journal{TBD}

\begin{document}

\begin{frontmatter}

\title{Linear/Quadratic Programming-Based Optimal Power Flow using Linear Power Flow and Absolute Loss Approximations}

\author[fen]{P. Fortenbacher \corref{cor1}}
\ead{fortenbacher@fen.ethz.ch}
\author[fen]{T. Demiray}
\ead{demirayt@fen.ethz.ch}

\cortext[cor1]{Corresponding author}

\address[fen]{Research Center for Energy Networks (FEN), ETH Zurich, Sonneggstrasse 28, 8092 Zurich }

\begin{abstract}
This paper presents novel methods to approximate the nonlinear AC optimal power flow (OPF) into tractable linear/quadratic programming (LP/QP) based OPF problems that can be used for power system planning and operation. We derive a linear power flow approximation and consider a convex reformulation of the power losses in the form of absolute value functions. We show four ways how to incorporate this approximation into LP/QP based OPF problems. In a comprehensive case study the usefulness of our OPF methods is analyzed and compared with an existing OPF relaxation and approximation method. As a result, the errors on voltage magnitudes and angles are reasonable, while obtaining near-optimal results for typical scenarios. We find that our methods reduce significantly the computational complexity compared to the nonlinear AC-OPF making them a good choice for planning purposes.
\end{abstract}

\begin{keyword}
Optimal Power Flow, Linear/Quadratic Programming, Power Flow Approximation
\end{keyword}

\end{frontmatter}

\section{Introduction}
\label{sec:intro}
\subsection{Motivation}
Optimal Power Flow (OPF) is indispensable for current research in power system operation and planning. OPF is widely used to find optimal expansion schemes \cite{Alguacil2003,DelaTorre2008,Taylor2011} for transmission networks in planning problems, or to find optimal generation schedules at operational level that minimize operational system costs. 

Especially for planning problems, it is crucial to have tractable formulations of multi-period OPF problems, since the incorporation of the nonlinear original OPF problem would impose a high computational burden. E.g. most advanced methods to solve such problems in the nonlinear AC-OPF framework are developed in \cite{Kourounis2018}. This is due to the intertemporal coupling of long investment horizons. Moreover, transmission planning methods need to incorporate a power flow approximation, since the combination of the nonlinear power flow equations and binary placement constraints makes the problem hard to solve. Therefore, current planning methods include either the well-known lossless DC power flow approximation \cite{Baringo2017}, a lossless approximation of the power flow in the full decision variable domain (active and reactive power, voltage magnitudes and angles) \cite{Taylor2011} or a power flow representation that does not operate in the full decision variable domain \cite{Alguacil2003,DelaTorre2008}. This can result in near-optimal or infeasible solutions, when neither network losses nor the full solution space are considered. 

Also unit commitment (UC) problems require power flow approximations and often binary decisions e.g. consideration of startup costs in the full decision variable space \cite{Murillo-Sannchez2013,Papavasiliou2013,Amjady2017}. This can be achieved by mixed-integer programming (MIP) frameworks that are either able to incorporate linear or semidefinite programming (SDP) power flow relaxations. However, relaxing planning or UC problems into a second order cone (SOC) programming problem \cite{Low2014} or into an SDP problem \cite{Molzahn2014} is still a complex optimization problem. Consequently, linear approximations are often the first choice to deal with the complexity issue. This also explains why UC problems are often divided into several stages \cite{Amjady2017} that reflect the binary decisions in the first stage, using a linear DC approximation, and then use the nonlinear OPF in the following stages at the cost of optimality, and/or computation time. 

In sum, there is still a clear need of linear OPF approximations that work in the full decision variable space and capture power losses. Hence, the objective of this paper is to find a linear and tractable approximation of the OPF problem in the full decision variable space of active/reactive power and voltage magnitudes/angles for universal grid topologies. 

\subsection{Related Work}
Finding reasonable linear power flow approximations for OPF problems is not a new research field. The first approaches included linearizing the power flow equations and passing this information to a Linear Programming (LP) solver. However, since this approximation does not hold for the entire operating range, the LP problem needs to be solved in an iterative way. Several papers \cite{Kirschen1988,Alsac1990,Olofsson1995} have used this solution approach, where they build the Jacobian of the power flow equations at a given operating point. The work featured in \cite{Coffrin2012,Coffrin2014,Koster2011,Bolognani2016} derives a linear approximation of the AC power flow equations, but does not show how these approximations can be incorporated into an OPF problem. The well-known DC-OPF is extended by a piecewise affine (PWA) loss model in \cite{Coffrin2012,Motto2002} to capture active power losses. However, these approaches only work in the active power domain. The authors of \cite{Yang2018,Zhang2013} suggested a linearized full OPF model with a power loss approximation. If the power losses are included as LP relaxations in the power balance constraints as done in \cite{Yang2018}, then the loss approximations will not be tight for negative locational prices (LMPs) on the power balance constraints, since fictitious losses would be generated that give the generation units more leeway to reduce the objective. The authors of \cite{Yang2018} cope with this issue by penalizing the active power losses in the objective, which can distort the objective value especially if the system is subject to heavy loading conditions. In contrast, \cite{Zhang2013} solves the fictitious losses problem by introducing an MIP PWA loss formulation for active and reactive power losses that is harder to solve as an LP problem. The authors of \cite{Zhang2013} also prove that an LP relaxation of the active power losses only holds for positive LMPs on the active power balance constraints if the reactive power losses are neglected. In \cite{Mhanna2016}, Mhanna et al. approximate the second order cone relaxations with linear relaxations resulting in a high number of linear constraints. Castillo et al. \cite{Castillo2016} use also an iterative approach to compute the optimal generator setpoints. The linear OPF method of~\cite{Horsch2018} does not capture losses and only operates in the decision variable domain of voltage angles and active power.
\begin{table}[t]
	\scriptsize
	\centering
	\setlength{\tabcolsep}{5pt}
	\caption{Comparison of suggested methods.}
	\label{tab:comp_methods}
	\begin{tabular}{lccccc}
		\hline
		Method & Problem & \multicolumn{2}{l}{Modeling of} & \multicolumn{2}{l}{Compatible with } \\
		&	&\multicolumn{2}{l}{Power Losses on} & \multicolumn{2}{l}{Negative LMPs on}\\
		&		& $p_\ell$ & $q_\ell$ & $p$ & $q$ \\
		\hline
		LOLIN-OPF & LP/QP & x & - & - & x \\
		LIN-OPF  &  LP/QP & - & - & x & x \\
		LINLOLIN-OPF & LP/QP & x & - & x & x \\
		MIP-OPF & MILP/MIQP & x & x & x&x \\
		\hline
	\end{tabular}	
\end{table}  
\subsection{Contribution}
The contribution of this paper is the development of novel tractable Linear/Quadratic Programming (LP/QP) based OPF methods that approximate the power flow over the entire operating range preventing us to iterate the problems. Our problems link the full decision variable domain with linear power flow approximations and capture the power losses by using absolute value loss approximations. Table~\ref{tab:comp_methods} summarizes the suggested methods listed by their names and their  capabilities indicated by the label (`x'). We developed four OPF methods, in which three of them are defined as LP/QP problems and one is formulated as an MILP/MIQP problem. The first LP/QP based method LOLIN-OPF includes the active power losses ${p}_\ell$ as epigraphs to the active power balance constraints, discards the reactive power losses ${q}_\ell$ and is only valid for non-negative LMPs on the active power balance constraints $p$. The second method LIN-OPF discards active and reactive power losses. The third method LINLOLIN-OPF is a combination of the first two aforementioned methods and considers active power losses. The fourth method MIP-OPF includes reactive and active power losses. The last three aforementioned methods can handle non-negative LMPs on reactive $q$ and active power balance constraints. Our methods have following characteristics:
\begin{enumerate}
	\item They can be solved by off-the-shelf solvers and the LP/QP methods are applicable for large grids.
	\item The presented approaches are universal to reflect any grid topology (meshed and radial) and any voltage levels (low voltage, distribution, and transmission grids).
	\item We showed by simulation that our approaches produce feasible AC solutions, if approximate solutions exist.
	\item The objective value error of our suggested methods is reasonable for typical test cases.   	
\end{enumerate}

The remainder of this paper is organized as follows. Section~\ref{sec:linapprox} derives the power flow approximation. Section~\ref{sec:approxOPF} shows how this approximation can be included into LP/QP based OPF formulations. Section~\ref{sec:results} analyzes the accuracy and optimality of our suggested OPF methods and Section~\ref{sec:conclusion} draws the conclusion.

\section{Linear Approximation}
\label{sec:linapprox}
We first derive the linear power flow approximations based on a two-bus example and extend this result to capture tap ratios, shunt elements, and line charging. Then, we introduce nodal admittance matrices to reflect any grid topology and size and incorporate this representation in several optimal power flow problems that have different features and are compliant with an LP/QP framework. 

\begin{figure}[t]
	\centering
	\def\svgwidth{0.8\columnwidth}
	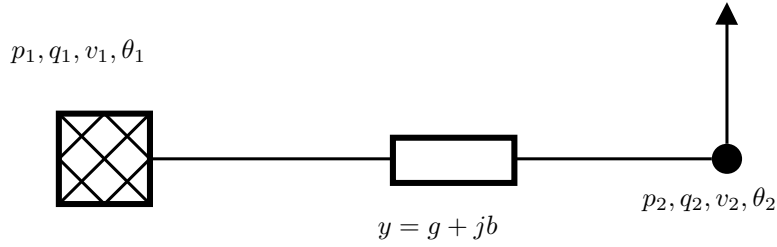
	\caption{Two-bus system to illustrate and derive the linear power flow and absolute loss approximations.}
	\label{fig:twobus}
\end{figure}

Based on Fig.~\ref{fig:twobus}, the nodal active $p_1,p_2$ and reactive $q_1,q_2$ powers are given by the nonlinear AC power flow equations that are for this case
\begin{equation}
\begin{array}{@{\hspace{2pt}}l@{\hspace{2pt}}c@{\hspace{2pt}}r@{\hspace{2pt}}r@{\hspace{2pt}}r@{\hspace{2pt}}}
p_1 & = & v_1^2 g &- v_1v_2 \cos(\theta_1 - \theta_2) g  & - v_1v_2 \sin(\theta_1-\theta_2) b \ , \\
p_2 &= &v_2^2 g  &- v_2v_1 \cos( \theta_2 - \theta_1) g & - v_2v_1 \sin(\theta_2-\theta_1) b \ , \\
q_1 & =& -v_1^2 b& + v_1v_2 \cos( \theta_1-\theta_2) b  &- v_1v_2\sin(\theta_1-\theta_2)g \ ,\\
q_2 & =& -v_2^2 b & + v_2v_1 \cos( \theta_2-\theta_1) b &- v_2v_1\sin(\theta_2-\theta_1)g \ ,
\end{array} \label{eq:non_pf_eq}
\end{equation}
\noindent where $v_1,v_2$ are the per unit nodal voltage magnitudes, $\theta_1,\theta_2$ are the voltage angles, $g$ is the per unit line conductance and $b$ the per unit line susceptance. 
\subsection{Absolute Loss Approximation}
\label{sec:approx}
Existing work focuses on power loss models that capture active power losses \cite{Martin2017,Motto2002,Zhang2013}. While \cite{Motto2002,Zhang2013} use PWA models representing linear constraints in voltage angles, Martin et al. showed the importance on incorporating voltage magnitudes to improve the approximation quality. However, these works consider transmission networks and need a higher number of linear constraints, which increases the problem complexity. In contrast, our power loss approximation captures active and reactive power losses and reduces the amount of constraints at the cost of optimality. Here, an absolute value formulation, which is derived in this section, constitutes the minimum size of  constraints.

The incurred active $p_\ell$ and reactive $q_\ell$ power losses for the two-bus system can be calculated by 
\begin{align}
p_\ell &= p_1 + p_2  = & (v_1^2 + v_2^2) g - 2v_1v_2\cos(\theta_1-\theta_2) g \ , \label{eq:exact_losses_p} \\
q_\ell &= q_1 + q_2  = & -(v_1^2 + v_2^2)b + 2v_1v_2\cos(\theta_1-\theta_2) b \ . \label{eq:exact_losses_q} 
\end{align}
\noindent  Let $v_1,v_2 = 1$ then we can find an absolute power loss approximation as a function of the voltage angle difference for active and reactive power as follows 
\begin{align}
p_\ell & = &2 (1-\cos(\theta_1-\theta_2))g &\approx  &  |\theta_1-\theta_2| 2 k_1 g \ , \label{eq:loss_theta_p}\\
q_\ell & = &-2 (1-\cos(\theta_1-\theta_2))b &\approx& -|\theta_1-\theta_2| 2 k_1 b \ , \label{eq:loss_theta_q}
\end{align}
\noindent where $k_1$ is a constant that represents the gradient of the absolute function associated with the voltage angle difference. Here, we approximate $(1-\cos(\theta_1-\theta_2))$ with $k_1|\theta_1-\theta_2|$. If we let $\theta_1-\theta_2 = 0$ then we obtain approximations in terms of absolute values that are functions of the voltage magnitude difference: 
\begin{align}
p_\ell & = &(v_1-v_2)^2 g &\approx & |v_1-v_2|2 k_2 g \ , \label{eq:loss_v_p}\\
q_\ell &= &-(v_1-v_2)^2 b &\approx & -|v_1-v_2|2 k_2 b  \ , \label{eq:loss_v_q}
\end{align}
\noindent where $k_2$ is a constant to approximate the losses associated with the voltage magnitude difference. Here, we approximate $(v_1-v_2)^2$ with $|v_1-v_2|2 k_2$.

By superposing the approximations~\eqref{eq:loss_theta_p} and \eqref{eq:loss_v_p} and superposing the approximations~\eqref{eq:loss_theta_q} and \eqref{eq:loss_v_q},  we approximate the active $p_\ell^\mathrm{approx}$ and reactive power losses $q_\ell^\mathrm{approx}$ as follows:
\begin{align}
p_\ell^\mathrm{approx} = &  \underbrace{|\theta_1-\theta_2|2 k_1g}_{2 p_\ell^{\Delta\theta}} + \underbrace{|v_1-v_2|2 k_2 g}_{2 p_\ell^{\Delta v}} \ , \label{eq:pl_approx} \\
q_\ell^\mathrm{approx} = & \underbrace{-|\theta_1-\theta_2|2 k_1 b}_{2 q_\ell^{\Delta\theta}} - \underbrace{|v_1-v_2|2 k_2 b}_{2 q_\ell^{\Delta v}} \ , \label{eq:ql_approx}
\end{align}
\noindent that are convex reformulations of the exact power losses \eqref{eq:exact_losses_p} and \eqref{eq:exact_losses_q}.

\subsection{Selection of $k_1,k_2$}
We have two degrees of freedom to approximate the power losses with the constants $k_1,k_2$. To parametrize those parameters we define the design parameters $\Delta \theta_\mathrm{d} = \theta_1-\theta_2, \Delta v_\mathrm{d} = v_1 - v_2$. They specify a usual voltage magnitude and angle difference between two nodes that are connected by a line. If we solve the equations \eqref{eq:loss_theta_p}, \eqref{eq:loss_theta_q} for $k_1$ and \eqref{eq:loss_v_p}, \eqref{eq:loss_v_q} for $k_2$, we obtain the following parametrizations for $k_1,k_2$

\begin{align}
k_1 & = \frac{1-\cos{\Delta \theta_\mathrm{d}}}{|\Delta\theta_\mathrm{d}|}  \approx \frac{|\Delta \theta_\mathrm{d}|}{2} \ , \\
k_2 & = \frac{|\Delta v_\mathrm{d}|}{2} \ .
\end{align}
\noindent The quality of the loss approximation depends strongly on the loading of the system and the selection of $k_1,k_2$. The errors can be reduced by evaluating the approximations around the operating points. Instead of optimizing $k_1,k_2$ for each specific load case, we choose typical static values for $\Delta \theta_\mathrm{d} = 0.05$ rad, $\Delta v_\mathrm{d} = 0.02$~p.u. by calculating the mean value of angular and magnitude differences obtained from typical load cases. This is different as proposed in \cite{Yang2018}, where a base case is needed to calculate grid specific loss factors. With this approach we weight the losses with respect to the susceptances and conductances. In this way it is possible to consider universal grid topologies ranging from low voltage grids usually having a high R/X ratio to transmission grids possessing a high X/R ratio. 
\begin{figure}
	\centering
	\subfloat[Active power losses]{%
		\includegraphics[width=0.7\columnwidth]{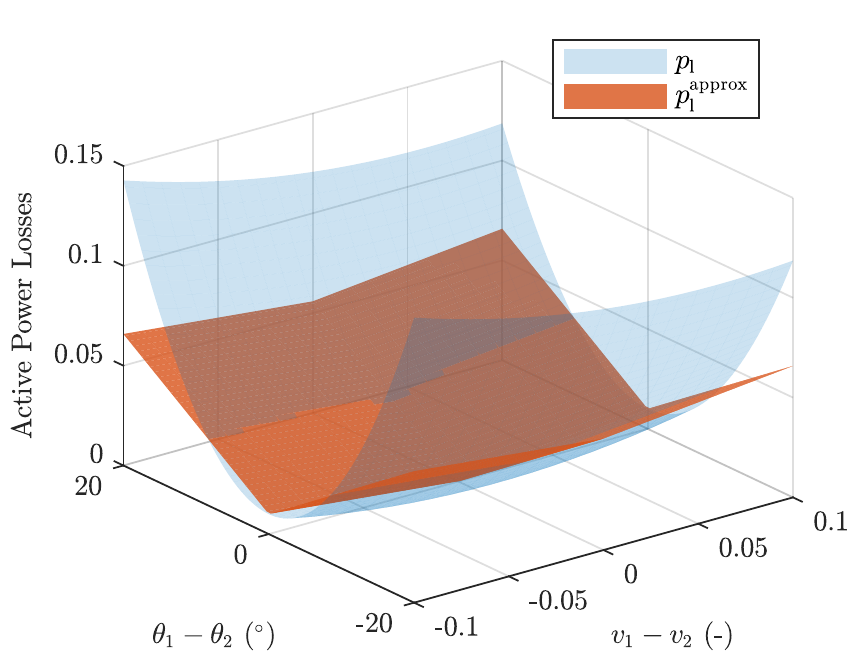} \label{fig:pl_approx}} \\
	\subfloat[Reactive power losses]{%
		\includegraphics[width=0.7\columnwidth]{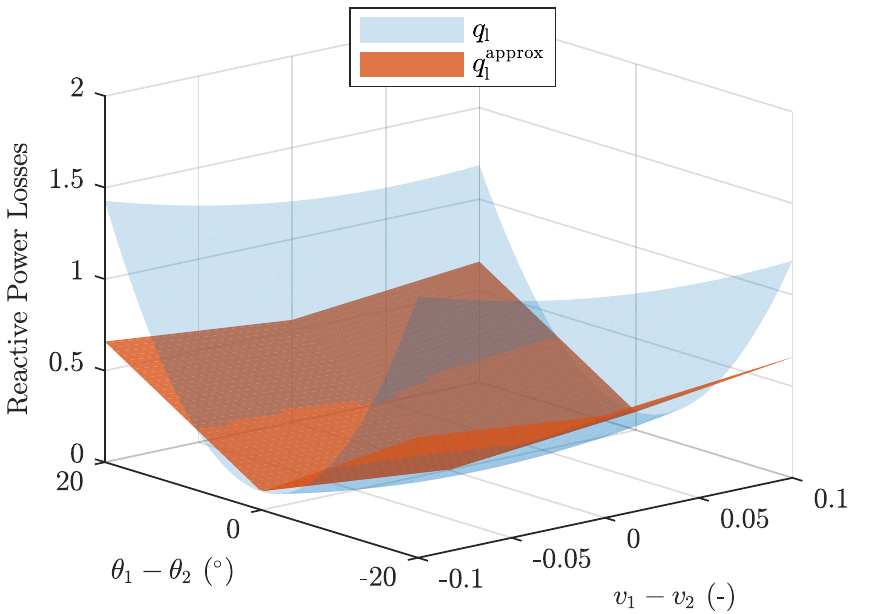} \label{fig:ql_approx}}
	\caption{Power loss comparison for the parameters $b=-10$ and $g=1$. The blue surface indicates the exact losses, while the red surfaces are the absolute power loss approximations.} 
	\label{fig:power_l_approx} 
\end{figure}
Figures~\ref{fig:pl_approx} and~\ref{fig:ql_approx} show the exact and the approximated power losses for the two-bus transmission system (Fig.~\ref{fig:twobus}) with the per unit susceptance $b=-10$ and the per unit conductance $g=1$. Note that this formulation is an approximation and not a relaxation, since there are loss regions that are underestimated above the values $\Delta \theta_\mathrm{d}$ and $\Delta v_\mathrm{d}$. This means that these errors translate to an underestimation of voltage angles and magnitudes. Hence, there is no guarantee that the approximated OPF solutions will lie inside the feasible original solution space.  
\subsection{Linear Power Flow Approximation}
\begin{figure}
	\centering
	\subfloat[Active power]{%
		\includegraphics[width=0.7\columnwidth]{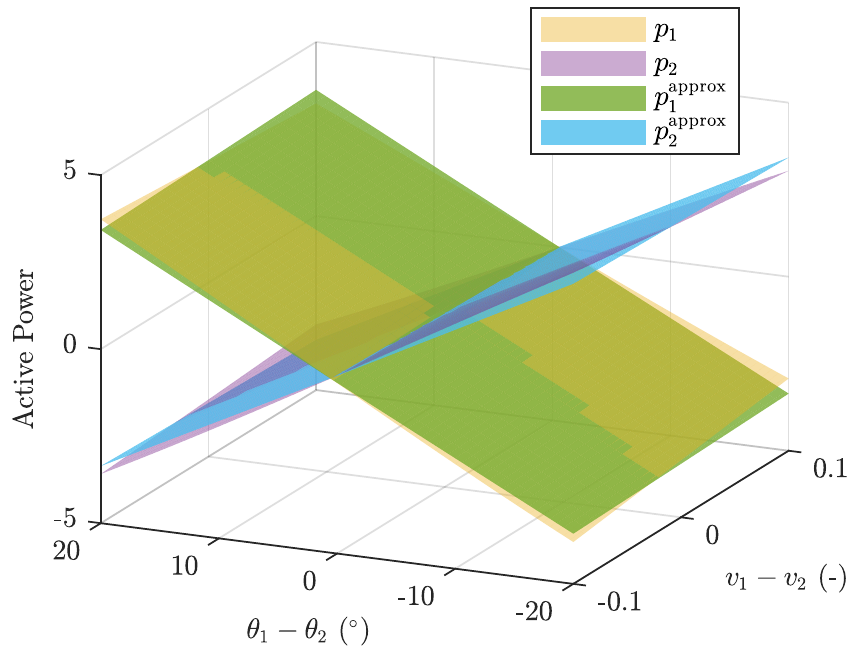} \label{fig:p_approx} }\\
	\subfloat[Reactive power]{%
		\includegraphics[width=0.7\columnwidth]{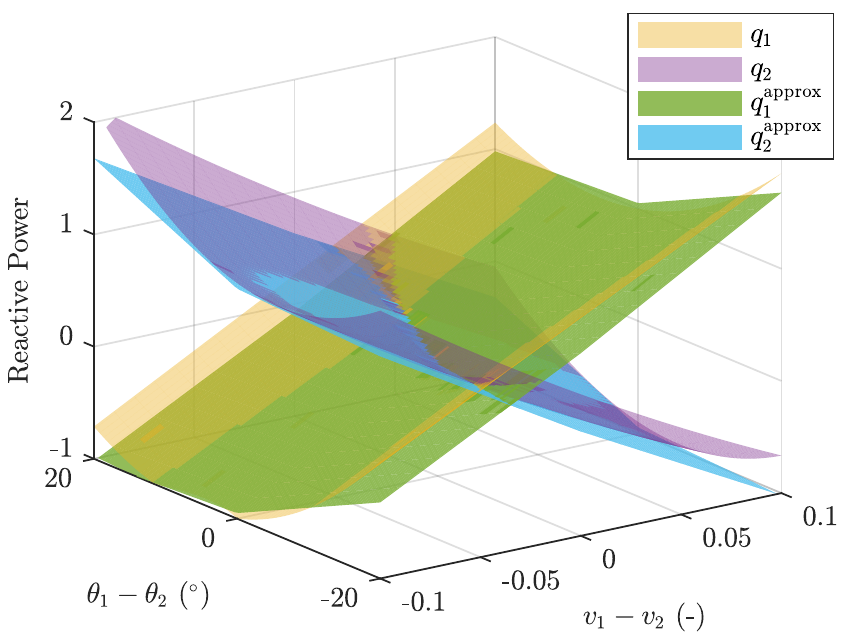} 	\label{fig:q_approx}}
	\caption{Power flow comparison. $p_1,p_2$ represent the nonlinear power flow equations \eqref{eq:non_pf_eq} and $p_1^\mathrm{approx},p_2^\mathrm{approx}$ are the convex power flow approximations \eqref{eq:power_approx}.} 
	\label{fig:power_approx} 
\end{figure}

We linearize the nonlinear power flow equations~\eqref{eq:non_pf_eq} by using the following approximations\footnote{Note that the DC power flow uses the same approximation \eqref{eq:dc} and the fast-decoupled power flow \cite{Stott1973} makes use of the approximation \eqref{eq:fast}.}:
\begin{align}
\cos(\theta_i-\theta_j) &\approx 1 \ , \label{eq:fast}\\
v_i^2-v_iv_j &\approx v_i-v_j \ , \\
v_i v_j\sin(\theta_i - \theta_j) & \approx(\theta_i - \theta_j) \ , \label{eq:dc}
\end{align}
to obtain
\begin{equation}
\begin{array}{@{\hspace{5pt}}l@{\hspace{5pt}}c@{\hspace{5pt}}r@{\hspace{5pt}}r@{\hspace{5pt}}r@{\hspace{5pt}}}
p_1 & \approx &(v_1-v_2) g &- (\theta_1-\theta_2) b  &+ p_\ell^{\Delta\theta}  + p_\ell^{\Delta v} \ , \\
p_2 & \approx &(v_2-v_1) g & - (\theta_2 - \theta_1) b&+ p_\ell^{\Delta\theta}  + p_\ell^{\Delta v} \ , \\
q_1 & \approx &-(v_1-v_2) b &- (\theta_1-\theta_2) g &+ q_\ell^{\Delta\theta}  + q_\ell^{\Delta v} \ , \\
q_2 & \approx &-(v_2-v_1) b& - (\theta_2 - \theta_1) g& + q_\ell^{\Delta\theta}  + q_\ell^{\Delta v} \ ,  \label{eq:power_approx}
\end{array}
\end{equation}
\noindent in which we also add the convex reformulations of the power losses $p_\ell^{\Delta v}$, $q_\ell^{\Delta v}$, $p_\ell^{\Delta\theta}$, $q_\ell^{\Delta\theta}$ derived from the previous Section~\ref{sec:approx}. As a result, the power flow approximations~\eqref{eq:power_approx} are convex, which can also be graphically verified in Figures~\ref{fig:p_approx} and \ref{fig:q_approx}. 
However, if these approximations are incorporated as power balance constraints in the OPF problem, then the nested absolute value functions \eqref{eq:pl_approx} and \eqref{eq:ql_approx} make the constraints nonconvex. For this reason, we relax the problem to a convex one by incorporating their epigraphs as linear constraints. 

The terms in $b$ for the nodal active powers $p_1,p_2$ correspond to the DC power flow approximation. All terms in $b$ correspond to the fast decoupled load flow expressions evaluated at the first iteration \cite{Stott1973}. However, the only difference is that we still have remaining expressions in $g$ that are crucial to capture power flows in distribution grids. The power flow approximations \eqref{eq:power_approx} follow the same structure as presented in \cite{Koster2011}, but are extended with the power losses. In contrast to \cite{Yang2018}, we do not use squared expressions for the voltage magnitudes.

\subsection{Extension for Line Charging, Transformer Tap Ratios and Shunts}
The aforementioned two-bus example considers only a series admittance $y$. In this section, we aim to extend our approach to incorporate line charging, transformer tap ratios and shunt elements. To capture these features, we use the standard $\pi$ branch model. The nodal admittance matrix $\vec{Y}_\mathrm{b}$ for the two-bus system is then e.g.
\begin{equation}
\vec{Y}_\mathrm{b} = 
\left[ \begin{array}{cc}
y_{ff}^i & y_{ft}^i \\
y_{tf}^i & y_{tt}^i
\end{array} \right] = \left[ \begin{array}{cc}
(y + j \frac{b_c}{2})\frac{1}{\tau^2} & -y\frac{1}{\tau \mathrm{e}^{-\theta_\mathrm{s}}} \\
-y \frac{1}{\tau \mathrm{e}^{\theta_\mathrm{s}}} & (y + j \frac{b_c}{2})
\end{array} \right]  , \label{ybus_2bus}
\end{equation}
\noindent where $y=g+jb$ is the complex series admittance, $\tau$ is the per unit tap ratio, $\theta_\mathrm{s}$ is the transformer shift angle, $b_\mathrm{c}$ is the per unit capacitive reactance of the line. We also define the matrix $\vec{Y}_\mathrm{b}'$, in which we only consider the series admittance with the complex tap ratios as follows 

\begin{equation}
\vec{Y}_\mathrm{b}' = \left[ \begin{array}{cc}
y_{ff}'^i & y_{ft}^i \\
y_{tf}^i & y_{tt}'^i
\end{array} \right] = \left[ \begin{array}{cc}
y\frac{1}{\tau \mathrm{e}^{-\theta_\mathrm{s}}} & -y\frac{1}{\tau \mathrm{e}^{-\theta_\mathrm{s}}} \\
-y\frac{1}{\tau \mathrm{e}^{\theta_\mathrm{s}}} & y\frac{1}{\tau \mathrm{e}^{\theta_\mathrm{s}}}
\end{array} \right] \quad. \label{ybus_2bus_star}
\end{equation}
This matrix is needed for the power flow approximation to correctly represent the power flow contribution on the voltage angles. 

In the same notation of \cite{matpower} we next generalize our approach to account for any grid topology. First, we define a grid that consists of $n_\mathrm{b}$ buses, $n_\mathrm{g}$ generators, and $n_\mathrm{l}$ lines. Any topology can be specified by constructing the node-branch incidence matrix $\vec{C}_\mathrm{br} \in \mathbb{Z}^{n_\mathrm{l} \times n_\mathrm{b}}$, from which we can derive the node-branch-from and node-branch-to incidence matrices $\vec{C}_\mathrm{f},\vec{C}_\mathrm{t} \in \mathbb{Z}^{n_\mathrm{l} \times n_\mathrm{b}}$. We introduce the generator active and reactive power injections $\vec{p}_\mathrm{g},\vec{q}_\mathrm{g} \in \mathbb{R}^{n_\mathrm{g} \times 1}$. The generator to bus mapping is specified with the matrix $\vec{C}_\mathrm{g} \in \mathbb{Z}^{n_\mathrm{g} \times n_\mathrm{b} }$. 

The branch-from and -to admittance matrices $\vec{Y}_\mathrm{f},\vec{Y}_\mathrm{t} \in \mathbb{C}^{n_\mathrm{l}\times n_\mathrm{b}}$ are calculated as
\begin{align}
\vec{Y}_\mathrm{f}   & = \text{diag}\{y_{ff}^1,\dots,y_{ff}^{n_\mathrm{l}}\}\vec{C}_\mathrm{f} + \text{diag}\{y_{ft}^1,\dots,y_{ft}^{n_\mathrm{l}}\}\vec{C}_\mathrm{t} \ , \\
\vec{Y}_\mathrm{t} & = \text{diag}\{y_{tf}^1,\dots,y_{tf}^{n_\mathrm{l}}\}\vec{C}_\mathrm{f} + \text{diag}\{y_{tt}^1,\dots,y_{tt}^{n_\mathrm{l}}\}\vec{C}_\mathrm{t} \ .
\end{align}
In the same straightforward way the adjusted versions of the branch admittance matrices $\vec{Y}'_\mathrm{f},\vec{Y}'_\mathrm{t} \in \mathbb{C}^{n_\mathrm{l}\times n_\mathrm{b}}$ are 
\begin{align}
\vec{Y}_\mathrm{f}'   & = \text{diag}\{y_{ff}'^1,\dots,y_{ff}'^{n_\mathrm{l}}\}\vec{C}_\mathrm{f} + \text{diag}\{y_{ft}^1,\dots,y_{ft}^{n_\mathrm{l}}\}\vec{C}_\mathrm{t}  ,\\
\vec{Y}_\mathrm{t}' & = \text{diag}\{y_{tf}^1,\dots,y_{tf}^{n_\mathrm{l}}\}\vec{C}_\mathrm{f} + \text{diag}\{y_{tt}'^1,\dots,y_{tt}'^{n_\mathrm{l}}\}\vec{C}_\mathrm{t} \ ,
\end{align}
in which we neglect the shunt admittances and line capacitances.

The nodal admittance matrix $\vec{Y}_\mathrm{b} \in \mathbb{C}^{n_\mathrm{b} \times n_\mathrm{b}}$, and its adjusted version $\vec{Y}_\mathrm{b}' \in \mathbb{C}^{n_\mathrm{b} \times n_\mathrm{b}}$ are determined by
\begin{align}
\vec{Y}_\mathrm{b} & = \vec{C}_\mathrm{f}^T \vec{Y}_\mathrm{f} + \vec{C}_\mathrm{t}^T\vec{Y}_\mathrm{t} + \text{diag}\{y_\mathrm{sh}^1,\dots, y_\mathrm{sh}^{n_b} \} \ , \\
\vec{Y}_\mathrm{b}' & = \vec{C}_\mathrm{f}^T \vec{Y}'_\mathrm{f} + \vec{C}_\mathrm{t}^T\vec{Y}'_\mathrm{t}  \ , 
\end{align}  
where $y_\mathrm{sh}^i$ are the per unit shunt admittances.

We extend the decision variables $\vec{\theta},\vec{v} \in \mathbb{R}^{n_\mathrm{b} \times 1},\vec{p}_\ell^{\Delta \theta},\vec{p}_\ell^{\Delta v},\vec{q}_\ell^{\Delta\theta},\vec{q}_\ell^{\Delta v} \in \mathbb{R}^{n_\mathrm{l} \times 1} $ to reflect any grid topology. The nodal active and reactive power injections $\vec{p},\vec{q} \in \mathbb{R}^{n_\mathrm{b} \times 1}$ are split into
\begin{align}
\vec{p} = \vec{C}_\mathrm{g}\vec{p}_\mathrm{g} - \vec{p}_\mathrm{d} \ ,\\
\vec{q} = \vec{C}_\mathrm{g}\vec{q}_\mathrm{g} - \vec{q}_\mathrm{d} \ ,
\end{align}
\noindent where $\vec{p}_\mathrm{d},\vec{q}_\mathrm{d}\in \mathbb{R}^{n_\mathrm{b} \times 1} $ are the active and reactive load vectors.

Under these definitions, we can find a more general matrix representation for the active power flow approximation as
\begin{equation}
\left[\begin{array}{@{\hspace{0pt}}c@{\hspace{2pt}}|@{\hspace{2pt}}c@{\hspace{2pt}}|@{\hspace{2pt}}c@{\hspace{2pt}}|@{\hspace{2pt}}c@{\hspace{2pt}}|@{\hspace{2pt}}c@{\hspace{0pt}}} -\Im\{\vec{Y}_\mathrm{b}'\} & \Re\{\vec{Y}_\mathrm{b} \} & -\vec{C}_\mathrm{g} & |\vec{C}_\mathrm{br}|^T & |\vec{C}_\mathrm{br}|^T \end{array} \right] \left[\begin{array}{@{\hspace{1pt}}c@{\hspace{1pt}}} \vec{\theta} \\ \vec{v} \\ \vec{p}_\mathrm{g} \\  \vec{p}_\ell^{\Delta \theta} \\ \vec{p}_\ell^{\Delta v}\end{array}\right] = -\vec{p}_\mathrm{d} \label{eq:p_bal} ,
\end{equation}
\noindent  and for the reactive power flow as
\begin{equation}
\left[\begin{array}{@{\hspace{0pt}}c@{\hspace{2pt}}|@{\hspace{2pt}}c@{\hspace{2pt}}|@{\hspace{2pt}}c@{\hspace{2pt}}|@{\hspace{2pt}}c@{\hspace{2pt}}|@{\hspace{2pt}}c@{\hspace{0pt}}} -\Re\{\vec{Y}_\mathrm{b}'\} & -\Im\{\vec{Y}_\mathrm{b}\} & -\vec{C}_\mathrm{g} & |\vec{C}_\mathrm{br}|^T & |\vec{C}_\mathrm{br}|^T \end{array} \right] \left[\begin{array}{@{\hspace{1pt}}c@{\hspace{1pt}}} \vec{\theta} \\ \vec{v} \\ \vec{q}_\mathrm{g} \\  \vec{q}_\ell^{\Delta\theta} \\ \vec{q}_\ell^{\Delta v}\end{array}\right] = -\vec{q}_\mathrm{d}
\label{eq:q_bal} ,
\end{equation}
where $\Re,\Im$ denote the real and imaginary part of a complex number. Note that \eqref{eq:p_bal} and \eqref{eq:q_bal} capture the approximated power flow equations \eqref{eq:power_approx} for the two-bus system if \eqref{ybus_2bus} and \eqref{ybus_2bus_star} are inserted.

\subsection{Branch Flow Approximation}
To obtain tractable OPF problems, we linearly approximate the active and reactive power line flows at the from ends $\vec{p}_\mathrm{f},\vec{q}_\mathrm{f} \in \mathbb{R}^{n_\mathrm{l} \times 1}$ as
\begin{align}
\vec{p}_\mathrm{f} & =  -\Im\{\vec{Y}_\mathrm{f}'\}\vec{\theta} + \Re\{\vec{Y}_\mathrm{f}\}\vec{v} \ , \\
\vec{q}_\mathrm{f} & =  -\Re\{\vec{Y}_\mathrm{f}'\}\vec{\theta} - \Im\{\vec{Y}_\mathrm{f}\}\vec{v} \quad. 
\end{align}
\noindent Note that the power line losses are neglected by using this formulation. 

\section{Approximated Tractable Optimal Power Flow Problems}
\label{sec:approxOPF}
In this section we derive the formulations of the approximated OPF problems. First, we present a lossy LP/QP based OPF problem that incorporates the power flow approximations as linear constraints. Hence, we call this method LOLIN-OPF. Secondly, we define a lossless version of the OPF problem that we call LIN-OPF. By combining the first two methods, we develop a further consecutive method that can deal with negative LMPs on active and reactive power balance constraints and is called LINLOLIN-OPF. Thirdly, we provide a MIP formulation of the problem, where we include the power flow approximations as linear constraints with binary decision variables. This method we call MIP-OPF. 

\subsection{Lossy LP/QP based Optimal Power Flow (LOLIN-OPF) Problem}
\label{sec:linopf}
\begin{figure}
	\centering
	\def\svgwidth{0.5\columnwidth}
	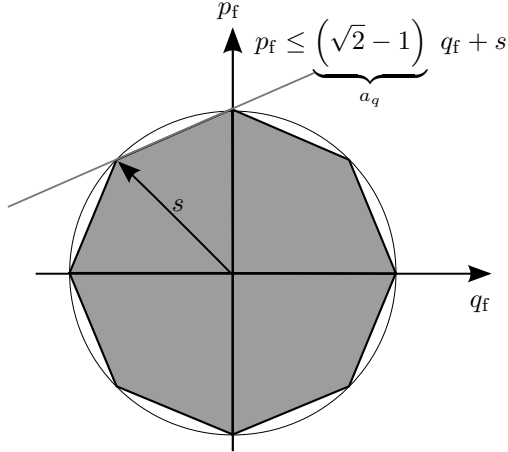
	\caption{Illustrative example for the convex line flow approximation for one power line. The gray line with the corresponding inequality specifies one segment of the circular PQ area.}
	\label{fig:pq_approx}
\end{figure}

With the introduced power flow approximations we can now formulate the OPF problem within a standard LP/QP framework. We specify the decision vector $\vec{x} = [\vec{\theta} \ \vec{v} \ \vec{p}_\mathrm{g} \ \vec{q}_\mathrm{g} \ \vec{p}_\ell^{\Delta \theta} \ \vec{p}_\ell^{\Delta v} \vec{q}_\ell^{\Delta \theta} \ \vec{q}_\ell^{\Delta v}]^T$. The objective of the OPF problem is to find the optimal active and reactive generator powers that minimize either a linear or quadratic cost objective. The approximated lossy LP/QP based Optimal Power Flow (LOLIN-OPF) problem  is
\begin{equation}
\begin{array}{@{\hspace{0pt}}l@{\hspace{6pt}}lll}
\text{LOLIN-OPF:}&\multicolumn{3}{@{\hspace{0pt}}l}{\displaystyle\min_{\vec{x}} f_p(\vec{p}_\mathrm{g}) + f_q(\vec{q}_\mathrm{g}) } \\
& \text{s.t.} & \eqref{eq:p_bal} , \eqref{eq:q_bal} \\   
&\text{(a)} &\multicolumn{2}{l}{ k_1\text{diag}\{\vec{g}\}\vec{C}_\mathrm{br} \vec{\theta}    - \vec{p}_\ell^{\Delta \theta} \leq \vec{0} }\\ 	
&\text{(b)} &\multicolumn{2}{l}{ -k_1\text{diag}\{\vec{g}\}\vec{C}_\mathrm{br} \vec{\theta}    - \vec{p}_\ell^{\Delta \theta} \leq \vec{0} }\\
&\text{(c)} &\multicolumn{2}{l}{ k_2\text{diag}\{\vec{g}\}\vec{C}_\mathrm{br}\vec{v} -  \vec{p}_\ell^{\Delta v} \leq \vec{0} } \\
&\text{(d)} &\multicolumn{2}{l}{ -k_2\text{diag}\{\vec{g}\}\vec{C}_\mathrm{br}\vec{v} -  \vec{p}_\ell^{\Delta v} \leq \vec{0} } \\
&\text{(e)} &\multicolumn{2}{l}{  k_1\text{diag}\{\vec{b}\}\vec{C}_\mathrm{br}\vec{\theta} -\vec{q}_\ell^{\Delta \theta} \leq \vec{0} }\\
&\text{(f)} &\multicolumn{2}{l}{ -k_1\text{diag}\{\vec{b}\}\vec{C}_\mathrm{br}\vec{\theta} -\vec{q}_\ell^{\Delta \theta} \leq \vec{0} }\\
&\text{(g)} &\multicolumn{2}{l}{  k_2\text{diag}\{\vec{b}\}\vec{C}_\mathrm{br} \vec{v} -\vec{q}_\ell^{\Delta v} \leq \vec{0} } \\
&\text{(h)} &\multicolumn{2}{l}{ - k_2\text{diag}\{\vec{b}\}\vec{C}_\mathrm{br} \vec{v} -\vec{q}_\ell^{\Delta v} \leq \vec{0}  } \\
&\text{(i)} & \multicolumn{2}{l}{-\vec{s} \leq \vec{p}_\mathrm{f} + \vec{A}_q \vec{q}_\mathrm{f} \leq \vec{s}} \\
&\text{(j)} & \multicolumn{2}{l}{-\vec{s} \leq \vec{p}_\mathrm{f} - \vec{A}_q \vec{q}_\mathrm{f} \leq \vec{s}} \\
&\text{(k)} & \multicolumn{2}{l}{-\vec{s} \leq \vec{A}_q\vec{p}_\mathrm{f} + \vec{q}_\mathrm{f}   \leq \vec{s}} \\ 
&\text{(l)} & \multicolumn{2}{l}{-\vec{s} \leq \vec{A}_q\vec{p}_\mathrm{f} - \vec{q}_\mathrm{f}   \leq \vec{s}} \ ,\\ 
\label{eq:lp}
\end{array}
\end{equation}
where $f_p,f_q$ are either linear or quadratic generator cost functions. The constraints \eqref{eq:p_bal}, \eqref{eq:q_bal} specify the nodal active and reactive power balance equations. The inequalities~\mbox{(\ref{eq:lp}a-h)} represent the epigraphs of the approximated absolute power loss functions derived in \eqref{eq:pl_approx} and \eqref{eq:ql_approx} and can be regarded as LP relaxations. It is noteworthy that the power loss vectors only lie on these hyperplanes if the Lagrange multipliers on the active and reactive power balance constraints \eqref{eq:p_bal} and \eqref{eq:q_bal} are non-negative. In other words the constraints~(\ref{eq:lp}a-h) need to be binding to achieve a physical meaningful solution. Negative multipliers (also called  LMPs) would lead to fictitious losses allowing to consume more active or reactive power to lower the objective value. Negative LMPs could occur in highly congested systems or for negative cost functions. Since we often have the case that generator cost functions are specified in the active power domain only, it is common to have negative LMPs on the reactive power balance constraints, which means that we can only find tractable approximations by neglecting reactive power losses. In addition, it was shown in \cite{Zhang2013} that the active loss LP relaxations only hold for non-negative LMPs on the active power balance, if the reactive power losses are neglected. Due to these reasons, we remove the reactive power loss variables $\vec{q}_\ell^{\Delta v},\vec{q}_\ell^{\Delta \theta}$ from the problem and discard the constraints~(\ref{eq:lp}e-h) at the cost of accuracy. To deal with negative LMPs on the active power balance constraints, we introduce a consecutive method that is described in Section~\ref{sec:iter}. To generally include active and reactive power losses, we will formulate an MIP problem that will be discussed in Section~\ref{sec:mip}. The constraints \mbox{(\ref{eq:lp}i-l)} define convex polygons to approximate the circular PQ capability area. The vector $\vec{s}\in \mathbb{R}^{n_\mathrm{l} \times 1}$ specifies the apparent power line limits and $\vec{A}_q = \text{diag} \{\vec{a}_q \in \mathbb{R}^{n_\mathrm{l} \times 1} \}$ represents the derivatives of the lines that form a convex polygon as shown in Fig.~\ref{fig:pq_approx}. Here, we consider 8 convex segments to approximate the circular PQ area. 

\subsection{Lossless LP/QP based Optimal Power Flow (LIN-OPF) Problem}
If the LOLIN-OPF problem returns negative LMPs on the active power balance and we would like to avoid solving an MIP problem, we can define a lossless version of the OPF problem at the cost of accuracy.
We eliminate the power loss vectors from \eqref{eq:p_bal} and  \eqref{eq:q_bal}, so that the decision vector is $\vec{x}' = [\vec{\theta} \ \vec{v} \ \vec{p}_\mathrm{g} \ \vec{q}_\mathrm{g} ]^T$. Then, the lossless LP/QP based Optimal Power Flow (LIN-OPF) problem is:
\begin{equation}
\begin{array}{@{\hspace{0pt}}l@{\hspace{2pt}}l@{\hspace{4pt}}l@{\hspace{0pt}}}
\text{LIN-OPF:} & \multicolumn{2}{@{\hspace{0pt}}l}{\displaystyle\min_{\vec{x}'} f_p(\vec{p}_\mathrm{g}) + f_q(\vec{q}_\mathrm{g}) } \\
& \text{s.t.}  & \text{(\ref{eq:lp}i-l)} \\ 
& \text{(a)} & -\Im\{\vec{Y}_\mathrm{b}'\}\vec{\theta} + \Re\{\vec{Y}_\mathrm{b} \}\vec{v}  -\vec{C}_\mathrm{g} \vec{p}_\mathrm{g} = -\vec{p}_\mathrm{d} \\
& \text{(b)} & -\Re\{\vec{Y}_\mathrm{b}'\}\vec{\theta} - \Im\{\vec{Y}_\mathrm{b} \}\vec{v}  -\vec{C}_\mathrm{g} \vec{q}_\mathrm{g} = -\vec{q}_\mathrm{d}  .
\label{eq:linOPF}
\end{array}
\end{equation}

\subsection{Consecutive lossy LP/QP based Optimal Power Flow (LINLOLIN-OPF) Problem}  
\label{sec:iter}
By solving the previous LIN-OPF method and running a second optimization problem it is possible to consider power losses for negative LMPs. For the second problem, we modify the LOLIN-OPF problem (\ref{eq:lp}) by fixing the power loss constraints with respect to the LIN-OPF solution. In particular, we equalize the power loss constraints (\ref{eq:lp}a-d) that are active in the directions of the voltage magnitude and angle differences obtained from the LIN-OPF solution and discard the remaining constraints being not active associated with the LIN-OPF solution.   

\subsection{Mixed Integer LP/QP based Optimal Power Flow \mbox{(MIP-OPF)} Problem}
\label{sec:mip}
As discussed in Section~\ref{sec:linopf} the optimal solution might not be binding with respect to the power loss constraints. An alternative but computationally more expensive formulation can be obtained by expressing the power loss constraints through an MIP approach. For this, we restructure the decision vector to $\vec{x}'' = [\vec{\theta} \ \vec{v} \ \vec{p}_\mathrm{g} \ \vec{q}_\mathrm{g} \ \Delta \vec{\theta} \ \Delta \vec{v} \  \vec{b}^{\theta} \ \vec{b}^v]^T$. The power loss vectors are now replaced by $ {\Delta\vec{\theta}},{\Delta\vec{v}} \in \mathbb{R}^{n_\mathrm{l} \times 1}$ specifying the absolute values of $|\theta_i-\theta_j|$ and $|v_i-v_j|$ in \eqref{eq:pl_approx} and \eqref{eq:ql_approx}. We introduce the binary variables $\vec{b}^{\theta},\vec{b}^v \in \mathbb{Z}^{n_\mathrm{l} \times 1}$ that are associated with ${\Delta\vec{\theta}},{\Delta\vec{v}}$. The approximated active and reactive power balance can be adjusted to comply with the new introduced decision vectors ${\Delta\vec{\theta}},{\Delta\vec{v}}$ as
\begin{equation}
\begin{array}{@{\hspace{2pt}}l@{\hspace{2pt}}}
	\left[\begin{array}{@{\hspace{5pt}}c@{\hspace{5pt}}c@{\hspace{5pt}}c@{\hspace{5pt}}c@{\hspace{5pt}}} -\Im\{\vec{Y}_\mathrm{b}'\} & \Re\{\vec{Y}_\mathrm{b} \} & -\vec{C}_\mathrm{g} &\vec{0}  \\
	-\Re\{\vec{Y}_\mathrm{b}'\} & -\Im\{\vec{Y}_\mathrm{b}\} & \vec{0} & -\vec{C}_\mathrm{g} 
	\end{array} \right. \\
	\left.\begin{array}{@{\hspace{2pt}}c@{\hspace{5pt}}c@{\hspace{2pt}}}
	 k_1\mathrm{diag} \{\vec{g} \}  |\vec{C}_\mathrm{br}|^T &  k_2\mathrm{diag} \{\vec{g} \}|\vec{C}_\mathrm{br}|^T \\
	 k_1\mathrm{diag} \{\vec{b} \} |\vec{C}_\mathrm{br}|^T & k_2\mathrm{diag} \{\vec{b} \} |\vec{C}_\mathrm{br}|^T
	\end{array} \right]
\end{array} \left[\begin{array}{@{\hspace{2pt}}c@{\hspace{2pt}}} \vec{\theta} \\ \vec{v}_\mathrm{m} \\ \vec{p}_\mathrm{g} \\ \vec{q}_\mathrm{g} \\ {\Delta\vec{\theta}} \\ {\Delta\vec{v}}\end{array}\right] = \left[\begin{array}{@{\hspace{2pt}}c@{\hspace{2pt}}}-\vec{p}_\mathrm{d} \\ -\vec{q}_\mathrm{d} \end{array} \right] \label{eq:mipbalance} .
\end{equation}
Then, the Mixed Integer LP/QP based Optimal Power Flow (MIP-OPF) problem is
\begin{equation}
\begin{array}{llll}
\text{MIP-OPF:}&\multicolumn{3}{l}{\displaystyle\min_{\vec{x}''} f_p(\vec{p}_\mathrm{g}) + f_q(\vec{q}_\mathrm{g})  } \\
&\text{s.t.} & \eqref{eq:mipbalance}, \text{(\ref{eq:lp}i-l)} \\   
&\text{(a)} &\multicolumn{2}{l}{-M(\vec{1}-\vec{b}^\theta) \leq -\vec{C}_\mathrm{br}\vec{\theta} \leq M\vec{b}^\theta } \\
&\text{(b)} &\multicolumn{2}{l}{\vec{0} \leq -\vec{C}_\mathrm{br}\vec{\theta} + \Delta\vec{\theta}\leq 2 M\vec{b}^\theta } \\
&\text{(c)} &\multicolumn{2}{l}{\vec{0} \leq  \vec{C}_\mathrm{br}\vec{\theta} + \Delta\vec{\theta}\leq 2 M(\vec{1}-\vec{b}^\theta) } \\
&\text{(d)} &\multicolumn{2}{l}{-M(\vec{1}-\vec{b}^v) \leq -\vec{C}_\mathrm{br}\vec{v} \leq M\vec{b}^v } \\
&\text{(e)} &\multicolumn{2}{l}{\vec{0} \leq -\vec{C}_\mathrm{br}\vec{v} + \Delta\vec{v}\leq 2 M\vec{b}^v } \\
&\text{(f)} &\multicolumn{2}{l}{\vec{0} \leq  \vec{C}_\mathrm{br}\vec{v} + \Delta\vec{v}\leq 2 M(\vec{1}-\vec{b}^v) \, ,}
\label{eq:mip}
\end{array}
\end{equation}
\noindent where the constraints~(\ref{eq:mip}a-f) specify a big M formulation of the absolute value functions in \eqref{eq:pl_approx} and \eqref{eq:ql_approx}. The variable $M$ has a considerable influence on the feasibility of the problem. It needs to be chosen sufficiently large to approximate the real (practical) range of the absolute values. Too large values might result in weak relaxations leading to branching the problem and hence to an increased computation time.

\section{Results}
\label{sec:results}
In this section we aim to show the performance of our proposed OPF methods compared to the nonlinear AC-OPF, the DC-OPF, and to existing approximation and relaxation OPF methods. In particular, we chose the quadratically-constrained (QC) OPF relaxation from \cite{Hijazi2017} and the lossy DC-OPF approximation from \cite{Coffrin2012}. We test our methods based on testcases from MATPOWER \cite{matpower} and from the benchmark library PGlib \cite{coffrin_pglib}.

\subsection{Implementation}
The hardware environment on which our algorithms run is equipped with an Intel Core i7-6600 processor. We implemented the suggested OPF methods within the MATPOWER framework \cite{matpower} and use the GUROBI~\cite{gurobi} solver parametrized with the barrier solver on four cores for solving our methods. The DC-OPF is also solved with GUROBI and with same solver settings. For all other methods, we use the interior point solver IPOPT~\cite{Wächter2006} with the integrated linear system solver MUMPS~\cite{mumps}. The approximation and relaxation methods are solved with PowerModels.jl~\cite{Coffrin2017}, from which we also adopt the names DCPLL for the lossy DC-OPF approximation and QCWR for QC-OPF relaxation. Both methods are formulated as QC programming (QCP) problems. We set the optimality tolerance for all solvers to 1e-6.

\subsection{Error Metrics}
\subsubsection{Voltage Errors}
We compare the voltage angles and magnitudes ($\vec{x}_\mathrm{s} = \vec{\theta},\vec{v}$) of the approximated OPF solutions with those from the exact power flow (PF) solutions ($\vec{x}_\mathrm{pf} = \vec{\theta}_\mathrm{pf},\vec{v}_\mathrm{pf}$). For the comparison, we set the PV buses in the PF solution according to the OPF solution and compute the root mean square (RMS) errors $\epsilon_{x_\mathrm{s}}$ on the nodal angle ($x_\mathrm{s}=\theta$) and magnitudes deviations ($x_\mathrm{s}=v$)

\begin{equation}
\epsilon_{x_\mathrm{s}} = \sqrt{\frac{(\vec{x}_{\mathrm{pf}} -\vec{x}_\mathrm{s} )^T(\vec{x}_{\mathrm{pf}} -\vec{x}_\mathrm{s} )}{n_\mathrm{b}} } \quad.
\end{equation}

Since the nodal errors propagate through the system, we are not able to identify the individual errors from the OPF approximations. The approximation error ($\epsilon_{{y}_\mathrm{s}}$) can be better captured by using the angular ($y_\mathrm{s} = \Delta\theta$) and magnitude ($y_\mathrm{s} = \Delta v$) differences along the lines and is defined as 
\begin{equation}
\epsilon_{y_\mathrm{s}} = \sqrt{\frac{(\vec{y}_{\mathrm{pf}} -\vec{y}_\mathrm{s} )^T(\vec{y}_{\mathrm{pf}} -\vec{y}_\mathrm{s} )}{n_\mathrm{l}} } \quad .
\end{equation}

\subsubsection{Objective Value Error}
We also define the objective value error between the AC-OPF solution and the approximated solutions. It is defined as
\begin{equation}
\epsilon_{\mathrm{O}} = \frac{f_{\mathrm{AC}}-f_z}{f_{\mathrm{AC}}} \quad,
\end{equation}
\noindent where $f_\mathrm{AC}$ is the objective value of the AC-OPF problem and $f_z$ is the objective value of the approximated problem $z$. 
   
\begin{table*}
\scriptsize
\setlength{\tabcolsep}{1pt}
\def\arraystretch{0.6}
        
\caption{OPF results for PGlib \cite{coffrin_pglib} and MATPOWER \cite{matpower} testcases including the bundled PEGASE \cite{Josz2016} test cases.}                                                                                           
\label{tab:OPFresults}             
\begin{tabular}{|c|c|c|c|c|c|c|c|c|c|c|c|c|}                     
\hline  
\# & Testcase &  Method  & Problem & Objective & $\epsilon_{\mathrm{O}}$ &  \multicolumn{3}{c}{Magnitude Error (p.u.)} & \multicolumn{3}{c}{Angle Error ($^\circ$)} & Runtime \\
   &	&		 &			& (\$/h)			& (\%)	&	$\epsilon_v$ &  $\epsilon_{\Delta v}$ & $\max|\epsilon_{\Delta v}|$  & $\epsilon_\theta$ & $\epsilon_{\Delta\theta}$ &  $\max|\epsilon_{\Delta\theta}|$  & (sec) \\
\hline  
\multicolumn{13}{|c|}{MATPOWER \cite{matpower} test cases} \\                        
\hline                                      
1 & case118 & AC-OPF & NLP & 1.297e+05 & - & - & - & - & - & - & - & 8.0e-01 \\
  		 &  & DC-OPF & QP & 1.259e+05 & 2.86 & n.p. & n.p. & n.p. & n.p. & n.p. & n.p. & 2.7e-01 \\
  		 &  & LIN-OPF & QP & 1.259e+05 & 2.86 & 0.002 & 0.002 & 0.009 & 1.89 & 0.35 & 1.48 & 2.0e-01 \\                                           
	     &  & LOLIN-OPF & QP & 1.296e+05 & 0.07 & 0.002 & 0.002 & 0.009 & 0.95 & 0.18 & 0.99 & 3.6e-01 \\                                          
 		 &  & MIP-OPF & MIQP & 1.296e+05 & 0.07 & 0.001 & 0.001 & 0.004 & 0.42 & 0.20 & 0.79 & 3.7e+00 \\                                          
 		 &  & DCPLL & QCP & 1.299e+05 & -0.15 & n.p. & n.p. & n.p. & n.p. & n.p. & n.p. & 1.7e+00 \\                                         
 		 &  & QCWR & QCP & 1.293e+05 & 0.25 & 0.001 & 0.001 & 0.007 & 5.12 & 1.21 & 4.76 & 7.7e-01 \\    
\hline
2 & case300 & AC-OPF & NLP & 7.197e+05 & - & - & - & - & - & - & - & 3.9e-01 \\                                                          
 		 &  & DC-OPF & QP & 7.063e+05 & 1.87 & n.p. & n.p. & n.p. & n.p. & n.p. & n.p. & 1.6e-01 \\                                                
  		 &  & LIN-OPF & QP & 7.063e+05 & 1.86 & 0.024 & 0.012 & 0.069 & 20.00 & 1.12 & 11.13 & 1.6e-01 \\                                          
  		 &  & LOLIN-OPF & QP & 7.180e+05 & 0.24 & 0.021 & 0.011 & 0.070 & 4.33 & 0.35 & 2.05 & 3.1e-01 \\                                          
  		 &  & MIP-OPF & MIQP & 7.181e+05 & 0.22 & 0.008 & 0.006 & 0.058 & 4.11 & 0.33 & 1.83 & 3.3e+01 \\                                          
  		 &  & DCPLL & QCP & 7.194e+05 & 0.04 & n.p. & n.p. & n.p. & n.p. & n.p. & n.p. & 1.2e-01 \\                                                
  		 &  & QCWR & QCP & 7.187e+05 & 0.15 & 0.007 & 0.007 & 0.093 & 3.38 & 1.60 & 10.20 & 2.4e+00 \\                                            
\hline                                          
3 & case1354pegase & AC-OPF & NLP & 7.407e+04 & - & - & - & - & - & - & - & 1.8e+00 \\                                                    
				&  & DC-OPF & LP & 7.306e+04 & 1.36 & n.p. & n.p. & n.p. & n.p. & n.p. & n.p. & 1.4e-01 \\                                                
				&  & LIN-OPF & LP & 7.306e+04 & 1.36 & 0.025 & 0.009 & 0.073 & 15.07 & 0.76 & 7.72 & 5.8e-01 \\                                           
				&  & LOLIN-OPF & LP & 7.475e+04 & -0.92 & 0.019 & 0.006 & 0.060 & 1.13 & 0.35 & 4.80 & 2.4e+00 \\                                         
				&  & DCPLL & QCP & 7.415e+04 & -0.11 & n.p. & n.p. & n.p. & n.p. & n.p. & n.p. & 8.5e-01 \\                                               
				&  & QCWR & QCP & 7.402e+04 & 0.07 & 0.002 & 0.001 & 0.011 & 4.63 & 0.82 & 5.77 & 2.5e+01 \\                             
\hline                 
4 & case33bw	 & AC-OPF & NLP & 7.835e+01 & - & - & - & - & - & - & - & 1.0e+00 \\
				 &  & DC-OPF & LP & 7.430e+01 & 5.17 & n.p. & n.p. & n.p. & n.p. & n.p. & n.p. & 2.7e-01 \\                                                
				 &  & LIN-OPF & LP & 7.430e+01 & 5.17 & 0.004 & 0.001 & 0.005 & 0.03 & 0.02 & 0.11 & 2.2e-01 \\                                          
				 &  & LOLIN-OPF & LP & 8.231e+01 & -5.05 & 0.001 & 0.001 & 0.002 & 0.11 & 0.04 & 0.13 & 5.0e-01 \\                                         
				 &  & MIP-OPF & MILP & 8.089e+01 & -3.24 & 0.001 & 0.000 & 0.001 & 0.03 & 0.02 & 0.11 & 5.9e-01 \\                                        
				 &  & DCPLL & QCP & 7.679e+01 & 2.00 & n.p. & n.p. & n.p. & n.p. & n.p. & n.p. & 1.8e+00 \\                                               
				 &  & QCWR & QCP & 7.835e+01 & 0.00 & 0.000 & 0.000 & 0.001 & 0.12 & 0.09 & 0.39 & 1.2e-01 \\                                             
\hline                                                                           \hline                                                         
5 & cigre \cite{Fortenbacher2016}	& AC-OPF & NLP & -2.700e+00 & - & - & - & - & - & - & - & 4.8e-01 \\                                                     
		&  	& DC-OPF & QP & -2.957e+00 & -9.52 & n.p. & n.p. & n.p. & n.p. & n.p. & n.p. & 2.1e-01 \\                                              
		&  & LIN-OPF & QP & -3.095e+00 & -14.63 & 0.003 & 0.000 & 0.001 & 0.10 & 0.01 & 0.05 & 1.5e-01 \\                                        
		&  & LOLIN-OPF & QP & -2.713e+00 & -0.47 & 0.001 & 0.000 & 0.001 & 0.03 & 0.01 & 0.03 & 2.0e-01 \\                                        
		&  & MIP-OPF & MIQP & -2.714e+00 & -0.52 & 0.001 & 0.000 & 0.001 & 0.01 & 0.00 & 0.01 & 4.8e-01 \\                                       
		&  & DCPLL & QCP & -2.692e+00 & 0.30 & n.p. & n.p. & n.p. & n.p. & n.p. & n.p. & 1.1e-02 \\                                               
		&  & QCWR & QCP & -2.700e+00 & 0.00 & 0.001 & 0.000 & 0.001 & 1.64 & 0.30 & 1.23 & 8.3e-02 \\
\hline      
\multicolumn{13}{|c|}{PGlib \cite{coffrin_pglib} test cases} \\
\hline
6 & case2869\_pegase & AC-OPF & NLP & 2.605e+06 & - & - & - & - & - & - & - & 8.0e+00 \\                                                  
  			&  & DC-OPF & LP & 2.501e+06 & 4.00 & n.p. & n.p. & n.p. & n.p. & n.p. & n.p. & 6.4e-01 \\                                                
  			&  & LIN-OPF & LP & 2.505e+06 & 3.84 & 0.020 & 0.006 & 0.061 & 35.65 & 0.78 & 12.05 & 3.7e+00 \\                                          
 			 &  & LINLOLIN-OPF & LP & 2.646e+06 & -1.59 & 0.020 & 0.006 & 0.056 & 3.37 & 0.32 & 3.77 & 1.2e+01 \\                                      
  			&  & DCPLL & QCP & 2.613e+06 & -0.29 & n.p. & n.p. & n.p. & n.p. & n.p. & n.p. & 3.2e+00 \\                                               
 			&  & QCWR & QCP & 2.577e+06 & 1.07 & 0.003 & 0.001 & 0.016 & 8.15 & 0.61 & 4.94 & 8.4e+01 \\      
\hline                                        
7 	& case9241\_pegase & AC-OPF & NLP & 6.775e+06 & - & - & - & - & - & - & - & 1.3e+03 \\                                                  
					&  & DC-OPF & LP & 6.541e+06 & 3.45 & n.p. & n.p. & n.p. & n.p. & n.p. & n.p. & 1.5e+00 \\                                                
					&  & LIN-OPF & LP & 6.510e+06 & 3.91 & 0.023 & 0.008 & 0.102 & 149.71 & 14.46 & 362.42 & 2.2e+01 \\                                       
					&  & LOLIN-OPF & LP & 7.315e+06 & -7.97 & 0.019 & 0.007 & 0.078 & 89.17 & 1.83 & 21.31 & 1.9e+02 \\                                       
					&  & DCPLL & QCP & 6.794e+06 & -0.28 & n.p. & n.p. & n.p. & n.p. & n.p. & n.p. & 1.6e+01 \\                                               
					&  & QCWR & QCP & 6.641e+06 & 1.98 & n.c. & n.c. & n.c. & n.c. & n.c. & n.c. & 7.5e+02 \\                                                
\hline
8 & case3120sp\_k\_\_sad & AC-OPF & NLP & 2.176e+06 & - & - & - & - & - & - & - & 6.9e+00 \\                                              
  						 &  & DC-OPF & LP & 2.252e+06 & -3.53 & n.p. & n.p. & n.p. & n.p. & n.p. & n.p. & 9.4e-01 \\
  						&  & LIN-OPF & LP & 2.160e+06 & 0.71 & 0.004 & 0.001 & 0.014 & 1.31 & 0.12 & 2.74 & 7.4e+00 \\
  						&  & LINLOLIN-OPF & LP & n.s. &  &  &  &  &  &  &  &  \\
  						&  & DCPLL & QCP & n.s. &  &  &  &  &  &  &  &  \\
  						&  & QCWR & QCP & 2.145e+06 & 1.41 & 0.003 & 0.001 & 0.033 & 1.15 & 0.24 & 2.49 & 6.3e+01 \\
\hline                                                                                                                                    
9 & case2383wp\_k\_\_api & AC-OPF & NLP & 2.791e+05 & - & - & - & - & - & - & - & 2.5e+00 \\                                              
  						&  & DC-OPF & LP & 2.791e+05 & 0.00 & n.p. & n.p. & n.p. & n.p. & n.p. & n.p. & 1.7e-01 \\
 						 &  & LIN-OPF & LP & 2.791e+05 & 0.00 & 0.002 & 0.001 & 0.008 & 1.49 & 0.11 & 1.03 & 1.8e+00 \\
  						&  & LINLOLIN-OPF & LP & 2.791e+05 & -0.00 & 0.001 & 0.001 & 0.010 & 3.62 & 0.22 & 2.35 & 3.2e+01 \\
  						&  & DCPLL & QCP & 2.791e+05 & 0.00 & n.p. & n.p. & n.p. & n.p. & n.p. & n.p. & 8.2e-01 \\
  						&  & QCWR & QCP & 2.791e+05 & 0.00 & 0.010 & 0.005 & 0.025 & 15.98 & 0.71 & 10.42 & 1.1e+01 \\
\hline                                                                                                                                    
10 & case2737sop\_k\_\_api & AC-OPF & NLP & 4.028e+05 & - & - & - & - & - & - & - & 4.1e+00 \\                                            
   						&  & DC-OPF & LP & 3.777e+05 & 6.24 & n.p. & n.p. & n.p. & n.p. & n.p. & n.p. & 2.5e-01 \\
   						&  & LIN-OPF & LP & 3.626e+05 & 9.98 & 0.003 & 0.001 & 0.010 & 2.51 & 0.18 & 4.99 & 1.2e+00 \\
   						&  & LINLOLIN-OPF & LP & 4.697e+05 & -16.59 & 0.003 & 0.001 & 0.009 & 3.50 & 0.21 & 3.22 & 1.8e+01 \\
  	 					&  & DCPLL & QCP & 4.062e+05 & -0.83 & n.p. & n.p. & n.p. & n.p. & n.p. & n.p. & 1.7e+00 \\
   						&  & QCWR & QCP & 3.626e+05 & 9.99 & 0.002 & 0.001 & 0.017 & 1.60 & 0.37 & 3.39 & 3.7e+01 \\
\hline
\end{tabular}
\end{table*} 

\subsection{OPF Comparison}
The comparison is performed on different test cases ranging from different grid topologies (meshed, radial), voltage levels (distribution, transmission systems), grid sizes, different loading conditions (typical, active power increase (api) conditions), and different operating conditions (small angular deviations (sad)). Table~\ref{tab:OPFresults} shows the OPF results for the performed test cases and different approximation approaches in terms of objective values, objective value errors, voltage errors, and computation times. The label 'n.p.` indicates that it is not possible to generate a PF solution for the DC-OPF and DCPLL-OPF, while 'n.c.' means that the PF does not converge. The label 'n.s.' indicates that no solution is found. In order to study the impact of negative LMPs on the active power balance, we always check in the AC-OPF solution on negative LMPs. If some exist (mainly in the api test cases), then we run the LINLOLIN-OPF instead of the LOLIN-OPF.
    
\subsubsection{Voltage Angle and Magnitude Errors}
\begin{figure}
	\centering
	\subfloat[Voltage angle comparison]{%
		\includegraphics[width = 0.8\columnwidth]{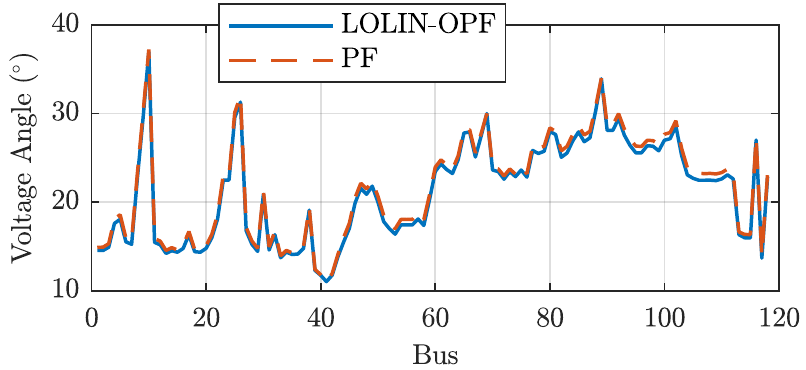}\label{fig:angle_error_118}} \\
	\subfloat[Voltage magnitude comparison]{%
		\includegraphics[width = 0.8\columnwidth]{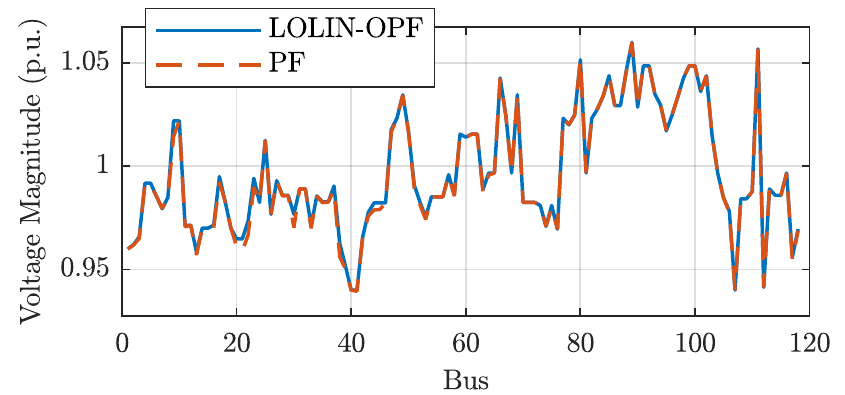}\label{fig:v_error_118}}
	\caption{LOLIN-OPF voltage magnitude and angle comparison with the power flow solution (PF) for the IEEE118 test grid.}
	\label{fig:comparision_118} 
\end{figure}

First, we aim to analyze the accuracy of our methods. As an example of results, Figures~\ref{fig:angle_error_118} and \ref{fig:v_error_118} show the angles and magnitudes for the LOLIN-OPF method and for the PF program for the IEEE~118 test grid. Although there is a small deviation in voltage angles, it can be observed that the curves match well. The angle offset can be explained by the fact that our OPF method overestimates the losses at the slack bus, so that this error propagates through the system. 

We show the angle and magnitude errors for all other test grids in Table~\ref{tab:OPFresults}. It can be observed that for our suggested methods a feasible PF solution exists, while for the QCWR relaxation for test \# 7 the power flow does not converge. 

Taking a closer look on the voltage magnitude errors, the MIP-OPF introduces the lowest. This is due to the fact that the MIP-OPF also considers the reactive power losses in the power flow approximation, while this is not the case for the LOLIN-OPF and LIN-OPF.   

We observe higher angle errors for the LIN-OPF method, which can be explained that the LIN-OPF version translates the underestimated active power generation setpoints to higher errors in voltage angles. Note that the DC-OPF solution would also generate such error. 
 
\subsubsection{Optimality}
Another fact that needs to be discussed is the optimality of our suggested OPF methods. Table~\ref{tab:OPFresults} lists the objective value errors ($\epsilon_\mathrm{O}$) of the analyzed methods. For the typical transmission test cases (\#1,2,3,6,7), we observe at maximum 3.9\% for the LIN-OPF, -1.59\% for the LINLOLIN-OPF, and -0.92\% for LOLIN-OPF. Note that we exclude the LOLIN-OPF result for case \#7, since our approach cannot consider negative resistances that correspond to an additional power feed-in. The negative values indicate that the associated methods overestimate the power losses in the grid to some extent. On average, the DC-OPF errors are higher than others. The DC-OPF and LIN-OPF solutions result almost in the same values and have lower objective values compared to the nonlinear OPF solution. This is due to the fact that these methods do not incorporate power losses. In contrast to the DC-OPF, the LIN-OPF also includes a voltage projection. The errors introduced by MIP-OPF and LOLIN-OPF are almost identical. 

The objective value errors for the distribution grid test cases \#4,5 are reasonable for the LOLIN-OPF (-5\%,-0.47\%). 

For extreme testcases (\#8,9,10) the objective value errors deviate (LINLOLIN-OPF -16\%, LIN-OPF 10\%) much more as for the typical test cases. This is due to the fact that a faithful approximation only holds for the given interval, in which $k_1$ and $k_2$ was selected for. While the LIN-OPF is feasible for all cases, the DCPLL-OPF and LINLOLIN-OPF do not converge for test case \#8. 

On average, the DCPLL and QCWR methods achieve lower objective value errors than our suggested methods in absolute terms. But at the same time it is noteworthy that the QCP problems are more complex problems than solving LP/QP problems.

\subsubsection{Complexity}
\begin{figure}
	\centering
	\includegraphics[width=0.8\columnwidth]{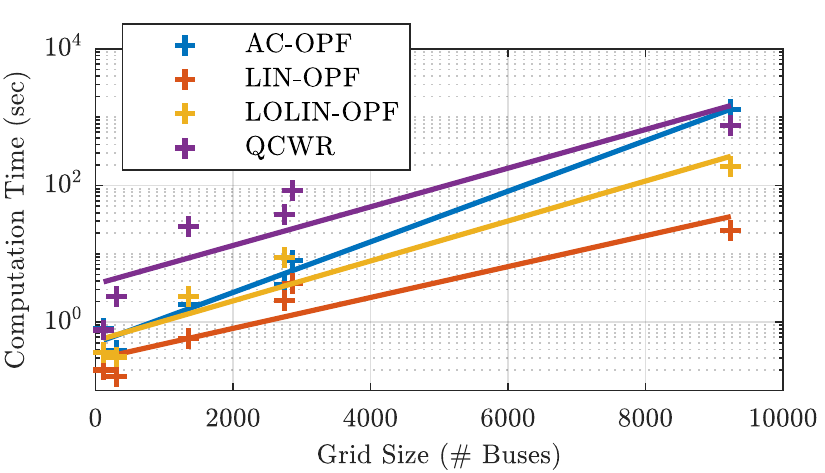}
	\caption{Computation time comparison between LOLIN-OPF, LIN-OPF, QCWR, and nonlinear AC-OPF and their corresponding lin-log regressions.}
	\label{fig:compTime}
\end{figure}

To assess the computational complexity of our methods, we consider only the typical test cases. We observe that the MIP-OPF does not scale well with respect to the grid size, such that this method is not applicable for large grids. The computation time results of the remaining methods that are in the full OPF domain are shown in Fig.~\ref{fig:compTime}. 

The lin-log regression of the QCWR method is almost shifted in parallel towards higher computation times in the directions of the LIN-OPF and LOLIN-OPF. This means that on average we achieve an improvement in computation time with the LIN-OPF method of one order of magnitude with respect to the QCWR relaxation. The LIN-OPF is almost five times faster than the LOLIN-OPF. Compared to the nonlinear AC-OPF the computation times of the LIN-OPF and LOLIN-OPF are much lower for larger grid sizes. Hence, it can be anticipated that this difference is even more pronounced for multi-period problems, where the grid size multiplies with the planning horizon.

\section{Conclusion}
\label{sec:conclusion}
In this paper we presented novel tractable OPF methods that work in the full decision domain of active/reactive power and voltage magnitudes/angles. We linearly approximate the power flow over the entire operating range avoiding the need to iterate the OPF problem. Our OPF methods can be used by efficient off-the-shelf LP/QP solvers. The obtained accuracy in terms of voltage magnitudes and angles is reasonable and we achieve near-optimal solutions for typical test scenarios and can reduce the computational complexity compared to the nonlinear \mbox{AC-OPF}.

\section*{Acknowledgment}
This project is carried out in the frame of the Swiss Centre for Competence in Energy Research on the Future Swiss Electrical Infrastructure (SCCER-FURIES) with the financial support of the Swiss Commission for Technology and Innovation (CTI - SCCER program).

\bibliography{literature}

\begin{thebibliography}{10}
\expandafter\ifx\csname url\endcsname\relax
  \def\url#1{\texttt{#1}}\fi
\expandafter\ifx\csname urlprefix\endcsname\relax\def\urlprefix{URL }\fi
\expandafter\ifx\csname href\endcsname\relax
  \def\href#1#2{#2} \def\path#1{#1}\fi

\bibitem{Alguacil2003}
N.~Alguacil, A.~L. Motto, A.~J. Conejo, {Transmission expansion planning: A
  mixed-integer LP approach}, IEEE Transactions on Power Systems 18~(3) (2003)
  1070--1077.
\newblock \href {https://doi.org/10.1109/TPWRS.2003.814891}
  {\path{doi:10.1109/TPWRS.2003.814891}}.

\bibitem{DelaTorre2008}
S.~de~la Torre, A.~J. Conejo, J.~Contreras, {Transmission expansion planning in
  electricity markets}, IEEE Transactions on Power Systems 23~(1) (2008)
  238--248.
\newblock \href {https://doi.org/10.1109/TPWRS.2007.913717}
  {\path{doi:10.1109/TPWRS.2007.913717}}.

\bibitem{Taylor2011}
J.~A. Taylor, F.~S. Hover, {Linear relaxations for transmission system
  planning}, IEEE Transactions on Power Systems 26~(4) (2011) 2533--2538.
\newblock \href {https://doi.org/10.1109/TPWRS.2011.2145395}
  {\path{doi:10.1109/TPWRS.2011.2145395}}.

\bibitem{Kourounis2018}
D.~Kourounis, A.~Fuchs, O.~Schenk, Towards the next generation of multiperiod
  optimal power flow solvers, IEEE Transactions on Power Systems (2018)
  1--1\href {https://doi.org/10.1109/TPWRS.2017.2789187}
  {\path{doi:10.1109/TPWRS.2017.2789187}}.

\bibitem{Baringo2017}
L.~Baringo, A.~Baringo, A stochastic adaptive robust optimization approach for
  the generation and transmission expansion planning, IEEE Transactions on
  Power Systems 8950~(c) (2017) 1--1.
\newblock \href {https://doi.org/10.1109/TPWRS.2017.2713486}
  {\path{doi:10.1109/TPWRS.2017.2713486}}.

\bibitem{Murillo-Sannchez2013}
C.~E. Murillo-Sanchez, R.~D. Zimmerman, C.~{Lindsay Anderson}, R.~J. Thomas,
  {Secure planning and operations of systems with stochastic sources, energy
  storage, and active demand}, IEEE Transactions on Smart Grid 4~(4) (2013)
  2220--2229.
\newblock \href {https://doi.org/10.1109/TSG.2013.2281001}
  {\path{doi:10.1109/TSG.2013.2281001}}.

\bibitem{Papavasiliou2013}
A.~Papavasiliou, S.~S. Oren, Multiarea stochastic unit commitment for high wind
  penetration in a transmission constrained network, Operations Research 61~(3)
  (2013) 578--592.
\newblock \href {https://doi.org/10.1287/opre.2013.1174}
  {\path{doi:10.1287/opre.2013.1174}}.

\bibitem{Amjady2017}
N.~Amjady, S.~Dehghan, A.~Attarha, A.~J. Conejo, Adaptive robust
  network-constrained {AC} unit commitment, IEEE Transactions on Power Systems
  32~(1) (2017) 672--683.
\newblock \href {https://doi.org/10.1109/TPWRS.2016.2562141}
  {\path{doi:10.1109/TPWRS.2016.2562141}}.

\bibitem{Low2014}
S.~H. Low, Convex relaxation of optimal power flow -- part i: Formulations and
  equivalence, IEEE Transactions on Control of Network Systems 1~(1) (2014)
  15--27.
\newblock \href {https://doi.org/10.1109/TCNS.2014.2309732}
  {\path{doi:10.1109/TCNS.2014.2309732}}.

\bibitem{Molzahn2014}
D.~K. Molzahn, I.~A. Hiskens, {Moment-based relaxation of the optimal power
  flow problem}, 2014 Power Systems Computation Conference (2014) 1--7\href
  {https://doi.org/10.1109/PSCC.2014.7038397}
  {\path{doi:10.1109/PSCC.2014.7038397}}.

\bibitem{Kirschen1988}
D.~Kirschen, H.~{Van Meeteren}, {MW/voltage control in a linear programming
  based optimal power flow}, IEEE Transactions on Power Systems 3~(2) (1988)
  481--489.
\newblock \href {https://doi.org/10.1109/59.192899}
  {\path{doi:10.1109/59.192899}}.

\bibitem{Alsac1990}
O.~Alsa{\c{c}}, J.~Bright, M.~Prais, B.~Stott, {Further developments in
  LP-based optimal power flow}, IEEE Transactions on Power Systems 5~(3) (1990)
  697--711.
\newblock \href {https://doi.org/10.1109/59.65896}
  {\path{doi:10.1109/59.65896}}.

\bibitem{Olofsson1995}
M.~Olofsson, G.~Andersson, L.~S{\"{o}}der, {Linear programming based optimal
  power flow using second order sensitivities}, IEEE Transactions on Power
  Systems 10~(3) (1995) 1691--1697.

\bibitem{Coffrin2012}
C.~Coffrin, P.~{Van Hentenryck}, R.~Bent, {Approximating line losses and
  apparent power in AC power flow linearizations}, IEEE Power and Energy
  Society General Meeting (2012) 1--8\href
  {https://doi.org/10.1109/PESGM.2012.6345342}
  {\path{doi:10.1109/PESGM.2012.6345342}}.

\bibitem{Coffrin2014}
C.~Coffrin, P.~V. Hentenryck, A linear-programming approximation of {AC} power
  flows, INFORMS Journal on Computing 26~(4) (2014) 718--734.
\newblock \href {https://doi.org/10.1287/ijoc.2014.0594}
  {\path{doi:10.1287/ijoc.2014.0594}}.

\bibitem{Koster2011}
A.~C.~A. Koster, S.~Lemkens, {Designing AC Power Grids Using Integer Linear
  Programming}, Network Optimization 6701 (2011) 478--483.
\newblock \href {https://doi.org/10.1007/978-3-642-21527-8_52}
  {\path{doi:10.1007/978-3-642-21527-8_52}}.

\bibitem{Bolognani2016}
S.~Bolognani, S.~Zampieri, {On the existence and linear approximation of the
  power flow solution in power distribution networks}, IEEE Transactions on
  Power Systems 31~(1) (2016) 163--172.
\newblock \href {http://arxiv.org/abs/1403.5031} {\path{arXiv:1403.5031}},
  \href {https://doi.org/10.1109/TPWRS.2015.2395452}
  {\path{doi:10.1109/TPWRS.2015.2395452}}.

\bibitem{Motto2002}
A.~L. Motto, E.~D. Galiana, {Network-Constrained Multiperiod Auction for a
  Pool-Based Electricity Market}, IEEE Power Engineering Review 22~(6) (2002)
  58.
\newblock \href {https://doi.org/10.1109/MPER.2002.4312289}
  {\path{doi:10.1109/MPER.2002.4312289}}.

\bibitem{Yang2018}
Z.~Yang, H.~Zhong, A.~Bose, T.~Zheng, Q.~Xia, C.~Kang, {A Linearized OPF Model
  with Reactive Power and Voltage Magnitude: A Pathway to Improve the MW-Only
  DC OPF}, IEEE Transactions on Power Systems 33~(2) (2018) 1734--1745.
\newblock \href {https://doi.org/10.1109/TPWRS.2017.2718551}
  {\path{doi:10.1109/TPWRS.2017.2718551}}.

\bibitem{Zhang2013}
H.~Zhang, V.~Vittal, G.~T. Heydt, J.~Quintero, {A relaxed AC optimal power flow
  model based on a Taylor series}, 2013 IEEE Innovative Smart Grid
  Technologies-Asia (ISGT Asia) (2013) 1--5\href
  {https://doi.org/10.1109/ISGT-Asia.2013.6698739}
  {\path{doi:10.1109/ISGT-Asia.2013.6698739}}.

\bibitem{Mhanna2016}
S.~Mhanna, G.~Verbi{\v{c}}, A.~C. Chapman, {Tight LP approximations for the
  optimal power flow problem}, 19th Power Systems Computation Conference, PSCC
  2016\href {http://arxiv.org/abs/1603.00773} {\path{arXiv:1603.00773}}, \href
  {https://doi.org/10.1109/PSCC.2016.7540937}
  {\path{doi:10.1109/PSCC.2016.7540937}}.

\bibitem{Castillo2016}
A.~Castillo, P.~Lipka, J.-P. Watson, S.~S. Oren, R.~P. O'Neill, A successive
  linear programming approach to solving the {IV-ACOPF}, IEEE Transactions on
  Power Systems 31~(4) (2016) 2752--2763.
\newblock \href {https://doi.org/10.1109/TPWRS.2015.2487042}
  {\path{doi:10.1109/TPWRS.2015.2487042}}.

\bibitem{Horsch2018}
J.~H{\"{o}}rsch, H.~Ronellenfitsch, D.~Witthaut, T.~Brown,
  \href{http://www.sciencedirect.com/science/article/pii/S0378779617305138}{Linear
  optimal power flow using cycle flows}, Electric Power Systems Research 158
  (2018) 126 -- 135.
\newblock \href {https://doi.org/https://doi.org/10.1016/j.epsr.2017.12.034}
  {\path{doi:https://doi.org/10.1016/j.epsr.2017.12.034}}.
\newline\urlprefix\url{http://www.sciencedirect.com/science/article/pii/S0378779617305138}

\bibitem{Martin2017}
J.~A. Martin, I.~A. Hiskens, {Generalized Line Loss Relaxation in Polar Voltage
  Coordinates}, IEEE Transactions on Power Systems 32~(3) (2017) 1980--1989.
\newblock \href {https://doi.org/10.1109/TPWRS.2016.2595763}
  {\path{doi:10.1109/TPWRS.2016.2595763}}.

\bibitem{Stott1973}
B.~Stott, O.~Alsac, {Fast Decoupled Load Flow}, IEEE Transactions on Power
  Apparatus and Systems PAS-93~(3) (1974) 859--869.
\newblock \href {https://doi.org/10.1109/TPAS.1974.293985}
  {\path{doi:10.1109/TPAS.1974.293985}}.

\bibitem{matpower}
R.~D. Zimmerman, C.~E. Murillo-Sanchez, R.~J. Thomas, {MATPOWER}: Steady-state
  operations, planning, and analysis tools for power systems research and
  education, IEEE Transactions on Power Systems 26~(1) (2011) 12--19.
\newblock \href {https://doi.org/10.1109/TPWRS.2010.2051168}
  {\path{doi:10.1109/TPWRS.2010.2051168}}.

\bibitem{Hijazi2017}
H.~Hijazi, C.~Coffrin, P.~V. Hentenryck, {Convex quadratic relaxations for
  mixed-integer nonlinear programs in power systems}, Mathematical Programming
  Computation 9~(3) (2017) 321--367.
\newblock \href {https://doi.org/10.1007/s12532-016-0112-z}
  {\path{doi:10.1007/s12532-016-0112-z}}.

\bibitem{coffrin_pglib}
{The {IEEE PES} task force on benchmarks for validation of emerging power
  system algorithms}, {PGLib} optimal power flow benchmarks,
  \url{https://github.com/power-grid-lib/pglib-opf}, accessed: 2018-04-01.

\bibitem{gurobi}
{Gurobi Optimization, Inc.}, \href{http://www.gurobi.com}{Gurobi optimizer
  reference manual} (2016).
\newline\urlprefix\url{http://www.gurobi.com}

\bibitem{Wächter2006}
A.~W{\"a}chter, L.~T. Biegler, On the implementation of an interior-point
  filter line-search algorithm for large-scale nonlinear programming,
  Mathematical Programming 106~(1) (2006) 25--57.
\newblock \href {https://doi.org/10.1007/s10107-004-0559-y}
  {\path{doi:10.1007/s10107-004-0559-y}}.

\bibitem{mumps}
P.~R. Amestoy, I.~S. Duff, J.-Y. L'Excellent, J.~Koster, A fully asynchronous
  multifrontal solver using distributed dynamic scheduling, SIAM Journal on
  Matrix Analysis and Applications 23~(1) (2001) 15--41.
\newblock \href {https://doi.org/10.1137/S0895479899358194}
  {\path{doi:10.1137/S0895479899358194}}.

\bibitem{Coffrin2017}
C.~Coffrin, R.~Bent, K.~Sundar, Y.~Ng, M.~Lubin,
  \href{http://arxiv.org/abs/1711.01728}{{PowerModels.jl: An Open-Source
  Framework for Exploring Power Flow Formulations}}\href
  {http://arxiv.org/abs/1711.01728} {\path{arXiv:1711.01728}}.
\newline\urlprefix\url{http://arxiv.org/abs/1711.01728}

\bibitem{Josz2016}
C.~Josz, S.~Fliscounakis, J.~Maeght, P.~Panciatici,
  \href{http://arxiv.org/abs/1603.01533}{{AC} power flow data in {MATPOWER} and
  {QCQP} format: {iTesla}, {RTE} snapshots, and {PEGASE}} (2016) 1--7\href
  {http://arxiv.org/abs/1603.01533} {\path{arXiv:1603.01533}}.
\newline\urlprefix\url{http://arxiv.org/abs/1603.01533}

\bibitem{Fortenbacher2016}
P.~Fortenbacher, M.~Zellner, G.~Andersson, Optimal sizing and placement of
  distributed storage in low voltage networks, in: 2016 Power Systems
  Computation Conference (PSCC), IEEE, 2016, pp. 1--7.
\newblock \href {http://arxiv.org/abs/1510310} {\path{arXiv:1510310}}, \href
  {https://doi.org/10.1109/PSCC.2016.7540850}
  {\path{doi:10.1109/PSCC.2016.7540850}}.

\end{thebibliography}

\end{document}